\documentclass[review]{elsarticle}

\usepackage[T1]{fontenc}
\usepackage[latin1]{inputenc}
\usepackage{graphicx}
\usepackage{amssymb}
\usepackage{verbatim}
\usepackage{epic,eepic}
\usepackage[english]{babel}
\usepackage{lmodern}
\usepackage{color,soul}
\usepackage{color,soul}
\usepackage{epsfig}
\usepackage{txfonts}
\usepackage{subfigure}
\usepackage{float}
\usepackage{wasysym}
\usepackage{url}
\usepackage{hyperref}
\usepackage{rotating}

\usepackage{hyperref}

\journal{NIM A}

\begin{document}
\begin{frontmatter}

\title{A new gas-based proton-recoil telescope for quasi-absolute neutron flux measurements
between $0.2$ and $2\,$MeV neutron energy }

\author[cenbg,b3]{P.~Marini\corref{mycorrespondingauthor}}
\cortext[mycorrespondingauthor]{Corresponding author}
\ead{paola.marini@cea.fr}
\author[cenbg]{L.~Mathieu}
\author[cenbg]{M.~A\"iche} 
\author[cenbg]{J.-L. Pedroza}
\author[cenbg]{T.~Chiron}
\author[ceaIrfu]{
P.~Baron}
\author[cenbg]{S.~Czajkowski} 
\author[cenbg]{F.~Druillole}
\author[cenbg]{P.~Hellmuth}
\author[cenbg]{B.~Jurado}
\author[cenbg]{A.~Rebii}
\author[cenbg]{I.~Tsekhanovich}

\address[cenbg]{CENBG, CNRS/IN2P3-Universit\'e de Bordeaux, Chemin du Solarium B.P. 120, F-33175 Gradignan, France}
\address[b3]{Present address: CEA, DAM, DIF, F-91297 Arpajon, France}
\address[ceaIrfu]{IRFU/SPhN, CEA-Saclay, Cedex F-91191 Gif-sur-Yvette, France}

\date{\today}

\begin{abstract}
Absolute measurements of neutron flux are an essential prerequisite of neutron-induced cross section measurements, neutron beam lines characterization and dosimetric investigations.
A new gaseous detector  has been developed for measurements of $0.2$ to $2\,$MeV  neutron flux based on proton-recoil process. 
The detector, consisting of two segmented ionization chambers read by Micromegas technology, has beed conceived to provide quasi-absolute neutron flux measurements with an accuracy of $\sim3\%$.
The 
gas pressure flexibility makes the telescope non sensitive to $\gamma$ and electrons background, and 
therefore advantageous over  semi-conductor materials as a neutron flux instrument. 
The  adjustable gas pressure and H-sample thickness,  the use of Micromegas technology and the tracking capabilities allows the detection of neutrons on a  large dynamical range and down to $200\,$keV with a good rejection of scattered neutron events and random background.
\end{abstract}

\end{frontmatter}


\section{Introduction}\label{Introduction}
The essential prerequisite of neutron-induced cross section measurements, neutron-beam lines characterisation and dosimetric investigations is the absolute measurements of neutron flux.
In particular,  
neutron-induced cross section measurements have  important applications in the development of Generation IV nuclear systems \cite{chadwick,Aliberti2006}, are 
relevant for the stellar nucleosynthesis processes \cite{arnould2007,rauscher2013} and  can improve our understanding of  different nuclear reaction processes.
Despite their importance in several fields and international efforts to precisely measure them, large uncertainties and discrepancies exist on these data, especially for the radioactive nuclei, the actinides being the example to give \cite{wpec26}.

The neutron  flux impinging on a sample is typically determined with respect to well-known fission cross sections of the standard reactions ($^{235}$U(n,f), $^{238}$U(n,f), $^{237}$Np(n,f)), thus introducing strong correlations between independent measurements based on the same standard. Moreover, the accuracy of these standards ranges from $0.5\%$ to $10\%$ \cite{hopkins,hale}, which represents an important source of uncertainty in neutron-induced  cross section measurements. 

 Independent and more accurate 
 measurements can be carried out with respect to the $^{1}$H(n,p) elastic scattering cross section, which is evaluated with a precision better than $0.5\%$ from $1\,$meV to at least $20\,$MeV \cite{hopkins,hale}. 
The principle of detectors based on this standard is to ``convert'' the neutron flux into a proton flux, and measure the proton recoil. 
 
 A variety of proton-recoil detectors have been designed for different purposes such as the diagnostic of hot plasma sources \cite{iguchi, kaneko} 
  or flux measurements of high energy \cite{dagendorf,schuhmacher} 
 or low energy \cite{wu1999} 
 neutrons. 
In the energy range of $1-70\,$MeV, silicon-based proton-recoil telescopes are the preferred method for absolute determination of monoenergetic neutron flux  \cite{bame57,ryves1976}. 
Multiple stage (stack of) detectors are used 
  to lower or eliminate the background due to unwanted reactions (for instance (n,p) and (n,$\alpha$) reactions above $6\,$MeV).\\
 However nowadays,  there is a need of extending precise  cross section measurements towards lower neutron energies, namely from 
 $1\,$MeV down to $100\,$keV,
for the development of Generation IV nuclear systems \cite{highprioritylist}.
The extension of the $\sigma$(n,p)-based method at these energies  
requires the detection of low energy protons ($\lesssim1\,$MeV) in a neutron and $\gamma$-dense environment, thus  a clean discrimination of the signal from the detector background, which is technically difficult.

Plastic scintillators using time-of-flight technique \cite{beyer}  and proton-recoil proportional counters \cite{skyrme1952,bennet1962,bennet1972} are typically used in this low energy region. 
However, scintillators light output is known to be non-linear, $^{12}$C(n,x) reactions must be discriminated from (n,p) reactions and the detector efficiency is difficult to calculate accurately 
 \cite{wu1999,daub,kovash2011,schmidt_d2002}. On the other hand, the use of  proportional counters allows to detect very low energy protons, since the recoil takes place in the sensitive volume of the detector. However, when using these detectors one must deal with  the 
dependence on the ionizing particle energy of the gas ionizing potential,  important wall effects and interferences from photons and carbon recoils \cite{skyrme1952,babut,NolteECJ}, thus relying on simulations for the neutron flux determination.
 
Silicon-based proton-recoil telescopes  remain, in principle, the best choice for neutron flux measurements. Indeed their response is determined exclusively by the number of H atoms in the radiator material and the differential H(n,p) scattering cross section. 
In addition, the simplicity of the detection system allows one to accurately determine the detection efficiency (better than $1\%$ \cite{kessedjian2012}) and to achieve a good control over the experimental conditions.
However, the detection of low energy protons 
 requires the use of a single-stage silicon detector, and therefore demands a good understanding of possible background sources, as well as of procedures to remove or avoid  them \cite{MariniNIMNoise}.
A pioneering use of surface-barrier silicon detectors down to $700\,$keV was successfully performed at the  Oak Ridge Electron Linear Accelerator time-of-flight facility \cite{morgan1978}.
 However, measurements at continuous beam facilities (Van de Graaff facility at JRC-Geel, AIFIRA \cite{sorieul2014}...) are  of paramount importance for the nuclear data community, since they are completely independent from those performed at time-of-flight facilities (nTOF \cite{nTOF2013}, GELINA Time-of-flight facility at JRC-Geel \cite{gelina}...).\\
Advantages and experimental challenges of using a single-stage $50\,\mu$m silicon-based proton-recoil detector (PRD) for flux measurements down to $500\,$keV neutron energy at a continuous neutron beam facility were investigated in Ref.\cite{MariniNIMNoise}.
The background sources in the so-obtained neutron flux measurement were identified and it was shown that the most
significant contributions arise from   scattered neutrons impinging on H-contaminated surrounding materials, and from Compton electrons scattered by low energy $\gamma$-rays produced by the neutron source \cite{mariniRadMeasDPR}. 
The ratio of the proton-signal to electron-background was shown to be strongly correlated to the silicon detector thickness, indicating that the latter should be adapted  to the proton range.
Proton-recoil detectors based on silicon detectors are therefore not suited for measurements on a large dynamical range below $1\,$MeV.
To overcome their lack of flexibility, we developed a double-stage Gas-based Proton-Recoil Telescope (GPRT) for neutron flux measurments at energies below $1\,$MeV. Indeed, in this case the gas pressure can be adjusted to match  the maximum proton range for the energy of interest.
The new type of GPRT, developed by the ACEN group at CENBG (Centre d'Etudes Nucl\'eaires de Bordeaux Gradignan), aims at fulfilling the following requirements: the detector
\begin{enumerate}
\item[i.] allows a precise ($\sim3\%$) quasi-absolute
 neutron flux  measurement. This requires  a well-defined  geometry to precisely determine the detector efficiency  better than $1$-$2\%$. 
The uncertainty on the neutron flux will then be determined mainly by the uncertainty on the number of $^{1}$H atoms  in the neutron-proton converter. 
 \item[ii.] is adapted for a neutron energy range from $0.2$ to at least $2\,$MeV. This requires a small noise level of the detector and the associated electronics. The primary electrons amplification should be strong enough to detect and separate from electronics noise the small amount of ionization produced in the gas by less-than-$200\,$keV protons, and the associated electronics should have an adequate dynamical range to accept signals without saturation up to $2\,$MeV.
\item[iii.] has a low sensitivity to background electrons. The effective thickness of the detector must be adjustable to match the maximum proton range for a given neutron energy and to minimize the thickness traversed by electrons.

\item[iv.] properly works in high neutron density environments (a few $10^{5}\,$n/s/cm$^{2}$). The detector and the associated electronics must be resistant to radiation damages.

\item[v.] allows for a direct vs. scattered neutron discrimination.
Protons generated by direct and scattered neutrons differ for their deposited energy and range in the gas, smaller for the scattered ones. Good tracking capabilities are therefore necessary for the detector to allow scattered-neutrons discrimination.
\end{enumerate}

The paper is organized as follows: in Secs.\ref{Gaseous proton-recoil detector design} and \ref{Electronics and data acquisition system} we present the design of the detector, and the associated electronics and data acquisition systems, respectively. Sec.\ref{Experimental tests} reports on the measured performances of the detector, while Sec.\ref{Conclusions} concludes and discusses  further planned measurements.

\section{Gaseous proton-recoil telescope design}\label{Gaseous proton-recoil detector design}
\begin{figure}[b!]
\centering
\includegraphics[width=0.95\columnwidth,clip]{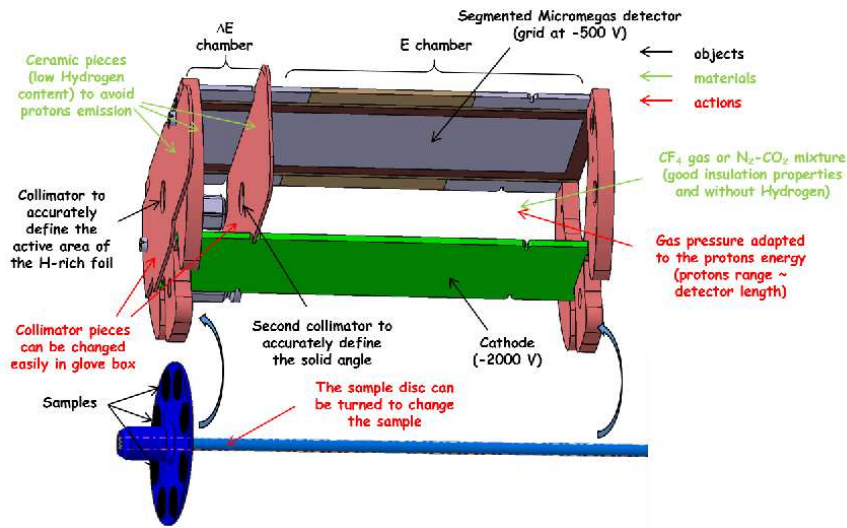}
\caption{(Color online) Schematic layout of the detector. The sample disk is shown separately for clarity purpose. The chamber is not shown.}
\label{fig:detector}       
\end{figure}

\begin{figure}[t!]
\centering
\includegraphics[width=0.95\columnwidth,clip]{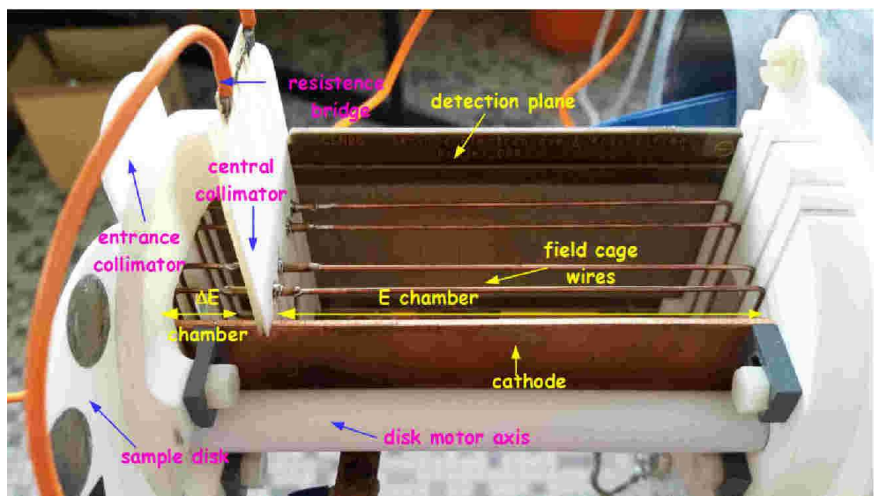} 
\caption{(Color online) Photo of the detector. Some of the components are indicated.}
\label{fig:photo-detector}      
\end{figure}

The lay-out of the GPR telescope described in this work is shown in Fig.\ref{fig:detector}, whereas a picture of the detector is presented in Fig.\ref{fig:photo-detector}.
The detector is constituted of a H-rich polypropylene foil,  mylar foil or a tristearine deposit, and a  segmented $\Delta$E-E ionization chamber, 
read by a 64 pads Micromegas-based detection plane. 
\paragraph{Detector and mounting structures} 
A special care was taken to build the detector and the dedicated chamber with a reduced amount of material,  to minimize the neutron scattering.
The  detector volume is limited by two parallel copper electrodes 
(4x12 cm$^{2}$), kept in place by two Macor ceramic pieces, which assure the positioning of the detector on the bottom of the chamber. The distance between the two electrodes is $4\,$cm.
Two $0.5\,$mm-thick Macor collimators well define the sample active area and the geometry of the detector. The use of Macor ceramic for all the insulating structures and collimators is motivated by the absence of $^{1}$H in the material, thus 
avoiding parasitic recoil-proton contribution.

\paragraph{Samples}
 Different $^{1}$H deposits 
 can be placed in six shallow circular holes drilled in a 
 Macor turning disk located $2\,$mm upstream the detection plane.   The thinnest samples could be of evaporated tristearine, while the thickest ones of polypropylene foils.
 An empty shallow circular hole is available for background measurements.
The step-by-step rotation of the disk is assured by a stepper motor, which allows one to select a sample  position without opening the chamber. 
 The possibility to select the sample thickness adapted to the neutron incident energy is crucial to obtain precise measurements on an extended dynamical energy range. 
 Indeed at low neutron energy the proton energy loss in the H-foil must be kept below $10\%$, to reduce protons energy spread \cite{TheseGrosjean}, thus requiring the use of very thin foils. 

A ceramic collimator, in the following referred to as entrance collimator, is placed $1.5\,$mm downstream the sample. This distance is a compromise between the possibility of rotating the sample disk and the need of a  well-defined  active area to precisely determine the amount of $^{1}$H in the converter.

\paragraph{Gas} The dedicated chamber containing the detector is filled with  gas at an adjustable pressure, from few tens of mbar to slightly above $1\,$bar.
 The pressure is regulated by a gas regulation unit and a piezoelectric pression sensor from Keller.

Two different hydrogen-free gases (CF$_{4}$ and N$_{2}-$CO$_{2}$) were used in order to prevent parasitic recoil-proton emission from the gas itself.
The two gases differ for their electron mobilities. The CF$_{4}$ has a 
high electron mobility, around $12\,$cm$/\mu$s between $20$ and $300\,$mbar and $1500\,$V, which allows 
to reduces the pile-up of the signals with a possible background noise.
On the contrary, the N$_{2}-$CO$_{2}$ electron mobility is  lower ($4\,$cm$/\mu$s), possibly limiting the counting rate, but allowing to measure the electrons drift time towards the anode. It should be noted that particle trajectory reconstructions in Time Projection Chambers is based on this method \cite{roger_actar}.

The adjustable gas pressure, from few tens to a few hundreds of mbar, allows one to adapt the maximum proton (or $\alpha$ particles, for tests with radioactive sources) range to the detector length,  to reduce as much as possible the detector sensitivity to Compton electrons \cite{MariniNIMNoise}.
Indeed electron energy loss in such a low density material is very small compared to that of protons, leading to a strongly reduced background.

\paragraph{Ionization chamber}
The ionization chamber is constituted by two parallel electrodes, generating an electric field perpendicular to the neutron beam direction. The anode is grounded, while the cathode is polarized at -$2000$V.
The $\Delta$E and E sides are $2\,$cm and $10\,$cm long, respectively.
A $0.5\,$mm thick ceramic collimator separates the two sides of the ionization chamber, 
providing a well-defined  geometry. 
In the following it is referred to as central collimator. A set of two collimators with different diameters were prepared. 
This allows one to limit the accepted proton recoil angles with respect to the  neutron beam direction and thus to limit the energy loss in the H-rich deposit also for the smallest incident energy.

 The coincidence between the $\Delta$E and E sides of the ionization chamber allows a first discrimination between direct and scattered neutrons to be performed. 

\begin{figure}[h!]
\centering
      \includegraphics[width=0.95\columnwidth,clip]{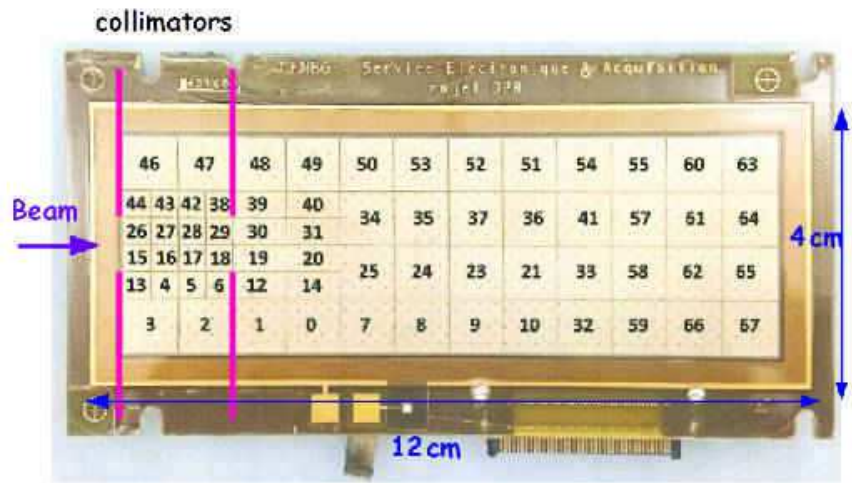} 
      \caption{Detection plane segmentation. The beam direction as well as the position of the collimators are indicated on the figure.}
\label{fig:detection plane}
\end{figure}

\paragraph{Detection plane}
The detection plane was made by CEA/Irfu. The plane, placed on the anode, is a $125\,\mu$m-gap Micromegas (MICRO-MEsh-GAseous Structure) detector \cite{giomataris,charpak}, segmented in $64$ pads.
The dimensions of the pads are of $10$x$10\,$mm$^{2}$, $5$x$10\,$mm$^{2}$ and $5$x$5\,$mm$^{2}$ 
depending on their position on the detection plane. A picture showing the pattern of pads is presented in Fig.\ref{fig:detection plane}.
The use of the Micromegas technology allows for the multiplication  factor of up to $10^{6}$ for the number of primary electrons generated by  proton ionization of the gas. Indeed, the number of primary electrons can be as low as $1000$ electrons per chamber side for $200\,$keV proton energy.
The detection plane segmentation allows both to reduce the electronic noise associated to the pad capacitance, and to perform a raw track analysis, to reject, at the trigger level or with software, accidental coincidences with abnormal trajectories.

\paragraph{Field Cage} To ensure the uniformity of the electric field among the electrods, the detector is equipped with two field cages, one for each side of the chamber (see Fig.\ref{fig:photo-detector}). Between and outside the two field cages there is the central collimator, which can be removed as needed.\\
Each field cage is made of four $1\,$mm-diameter copper wires streched around the perimeter of each side of the ionization chamber. The distance between the wires is $7\,$mm, except for the central ones, placed $12\,$mm away from each other not to mask the opening of the  collimator.
On the $\Delta$E side of the chamber, the wires are supported by a $0.8\,$mm-thick epoxy slab. Each wire is soldered into the slab to a copper electric track to close the circuit. 
Similarly, on the E-side of the chamber, $0.2\,$mm-thick wires are streched along the collimator to connect the two side of each wire and do not mask the collimator opening.\\
The electric supply is distributed via a resistence bridge soldered to the epoxy slab. A variable resistence, remotely controllable by users, serves for adjusting the resistence bridge to the selected grid and cathode potentials. 

\section{Electronics and data acquisition system}\label{Electronics and data acquisition system}
\begin{figure}[ht!]
      \includegraphics[width=0.95\columnwidth,clip]{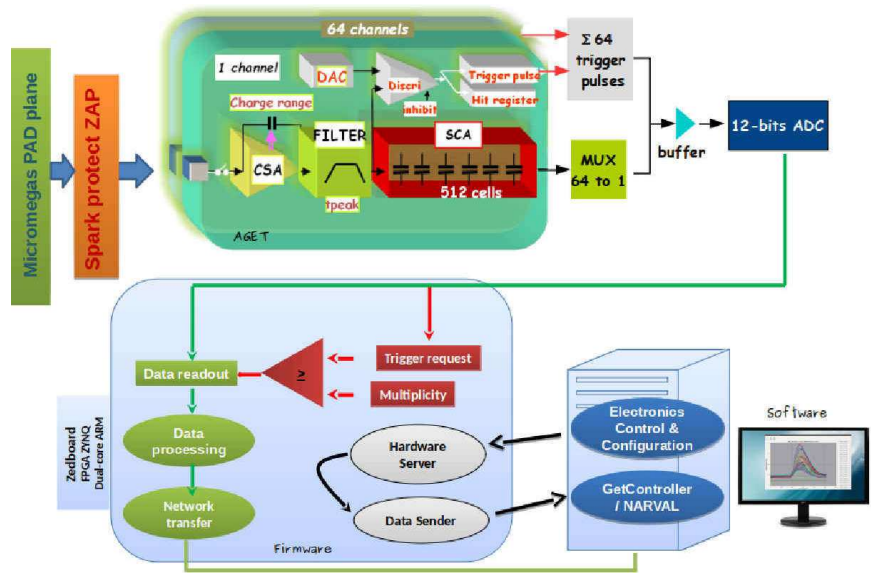} 
      \caption{The SAM electronics. The AGET and ADC modules are shown in the upper part of the sketch. The DAC module is not shown in the figure. The firmware is shown in the lower part of the sketch. Courtesy of J.~L. Pedroza. }
\label{fig:SAM electronics}
\end{figure}

The signal treatment is realised by the Single AGET Module (SAM) electronics and acquisition system developed from the Generic Electronic System for TPCs (GET) \cite{pollacco2012} by the CENBG electronics service. A 
schematic of the principle is shown in Fig.\ref{fig:SAM electronics}.
SAM consists of two main modules: a front-end module, based on the AGET circuit (ASIC for General Electronic for TPC) from the GET collaboration, and a ZedBoard commercial module with a Zynq XC7Z020 field-programmable gate array (FPGA) \cite{zedboard} .
\subsection{Front-end electronics}
\begin{figure}[t!]
      \includegraphics[width=0.95\columnwidth,clip]{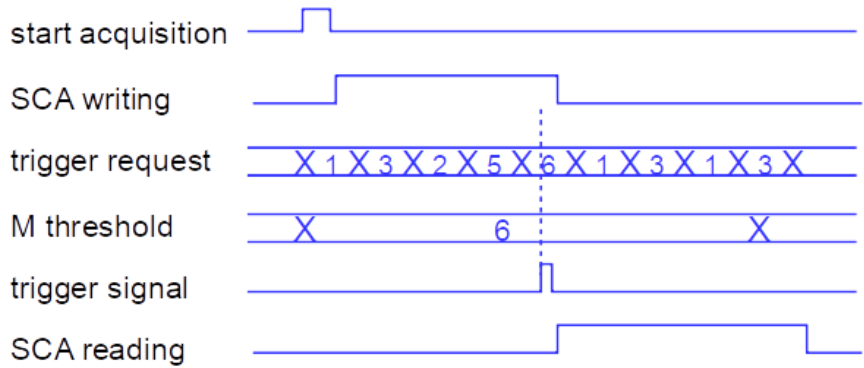} 
      \caption{Schema of the trigger request and data read-out time sequence. The multiplicity threshold set by users is labelled as ``M threshold'' in the figure. See text.
      }
\label{fig:chronogramme}
\end{figure}

Signals from the detection plane pads are fed into a spark protection circuit card (ZAP), through a High Speed Coaxial Cable Assembly from Samtec. The latters 
 serve as a hardware interface between the detector and the front-end board. 
Each channel of the ZAP card consists of a CR-filter followed by two ultra-fast push-pull low-noise diodes, limiting the high voltage excursion for each AGET channel input.\\
The SAM front-end module houses one 64-channels AGET chip, a 12-bits analog-to-digital converter (ADC), a charge injection generator and four inspection lines. 
Each AGET-channel  is constituted of a charge sensitive preamplifier (CSA) with four adjustable dynamical ranges (from $120\,$fC to $10\,$pC). 
To improve the signal-over-noise ratio, a shaper with $16$ adjustable peaking times (from $70\,$ns to $1\,\mu$s) filters the CSA output. The filtered signal is sampled in a 512-cells analog memory (SCA) with frequency from $1\,$MHz to $100\,$MHz, depending on the user needs. The filtered signal  is also sent to a discriminator to inform the system of the occurrence of an event. 
When the signal crosses the discriminator threshold, the output goes to a hit register, which records the number of the fired channel. A copy of the discriminator output signal (``trigger pulse'') is combined with the other 63 discriminator signals to create a multiplicity signal.
The latter 
can be obtained as the analog OR of $64$ or of $N$ user-selected discriminator outputs. The analog-multiplicity signal is coded by an external 12-bits ADC and is used as ``trigger request'' signal. 
The  ADC  operates in  continuous-read mode, at a frequency of $25\,$MHz, so that the information on the event multiplicity is continuously sent to the trigger module located in the firmware for comparison with the user-requested multiplicity and for possible trigger-signal generation. 
In addition, the continuous-read mode  allows to minimize the dead-time.\\
The 512-cells analog-memory data sampling is stopped by the trigger decision. In the read-out phase, the analog data from the 64 channels are multiplexed toward a single output and send to the 12-bits ADC at a readout frequency of $25\,$MHz.
To reduce the noise, two differential lines are used to send the stored signals to the ADC. Digitalized signals are then send to the AGET controller-block in the Zedboard.
A schema of the trigger request and data readout sequence is shown in Fig.\ref{fig:chronogramme}.

To test the system independently from the detector, the SAM board houses a pulser to emulate the charge injection through a set of capacitors. Thanks to a 14-bits current-mode digital-to-analog converter (AD9707), the SAM module can be tested and  calibrated without detector.
\subsection{Firmware}
SAM digital functions are  controlled from an embedded server in the Zynq FPGA installed on the ZedBoard. The server communicates to the different firmware modules via an Advanced eXtensible Interface AXI Bus.\\
The embedded system architecture of the Zedboard consists of four mains blocks linked to the server via the AXIS bus. The AGET controller-block manages the configuration of the chip and the acquisition of the analog signal from AGET, it processes trigger information to reduce the data size,  adds a time stamp for each event and prepares the data transfer through the TCP/IP link. A second block controls the ADC, sends frame and clock signals, and captures digital information from the ADC. A third block manages the charge injection pulse generator through the AD9707 digital-to-analog converter circuit. 
The last block controls the transmission of debug and environmental information through the inspection lines to the users.

The response time of the full system is of few ns, due to the time needed for the comparison between the multiplicity threshold and the multiplicity signal. Without a valild trigger signal each shaper output analog signal is stored and erased if not read out like a circular buffer. Finally there is no recovery time unless sever signal saturations occur (1 ms to evacuate the accumulated charges). \\
The readout can be performed on the basis of  hit channels or   user-selected channels 
or a full readout of the $64$ channels.
The dead-time of the electronics and acquisition system for a one channel readout is of 
$20.5\,\mu$s and for a full readout of the 64 channels is of $1.4\,$ms (64 channels x 512 capacitors at $25\,$MHz).
Depending on the TCP/IP and data transfer through a Zedboard direct memory access system (DMA), one could read events at, at least, $700\,$Hz.

\subsection{SAM DAQ software}
SAM is delivered with the GetController software to manage all parameters, visualize current signals and store data in binary file. The data format is based on GET Multiframe Metaformat model elaborated in the context of TPC instruments. The main improvement of the MultiFrame metaformat is the possibility 
to define binary formats for data
acquisition and serialization that are self-contained, layered, adapted to network transfers and evolving \cite{MFMdataFormat}. 
\section{Experimental tests}\label{Experimental tests}
The detector prototype was tested with quasi-monoenergetic  neutron beams at the AIFIRA facility of the CENBG  laboratory \cite{sorieul2014}. 
Fast neutrons with energies  from $200\,$keV to $1\,$MeV were produced in the $^{7}$Li(p,n) reaction, with protons accelerated to energies ranging from $2$ to $2.7\,$MeV. For every proton beam energy, the obtained neutron beam was monoenergetic at 0$^{\circ}$.\\
The detector response was studied under different experimental conditions (applied voltage, type of gas, gas pressure). 
Tests were also performed with and without the field cage. 
H-rich foils, with thickness of $4$ and $10\,\mu$m, were mounted on the sample disk.
A radioactive $3\,\alpha$ source ($^{233}$Pa, $^{241}$Am and $^{244}$Cm) of about $1\,$kBq  
 was also mounted on the sample disk for test and calibration purposes.
 The detector prototype was placed in its aluminum chamber filled with gas. The gas pressure was varied from $40$ to $400\,$mbar, depending on the used gas (CF$_{4}$ or N$_{2}-$CO$_{2}$) and on the neutron energy. The voltage applied to the Micromegas 
 was  the nominal -$500\,$V unless differently specified (see Sec.\ref{sec:electrostatic behavior}). The  voltage applied to the cathode was adapted consequently. The gas pressure was set to match the maximum proton range to the detector dimensions, so that the most energetic proton track ends in the previous-to-last pad of the detector. 
The H-rich sample was placed at a polar angle of $0^{\circ}$  with respect to the beam axis at a distance of a few cm from the neutron source (LiF target).  
Signals from all the detection plane pads were simultaneously read.
The acquisition trigger was built requiring a given multiplicity of fired detection pads. 
The GetControler software allowed us to visualize, event by event, the signals detected on each pad.
\begin{figure}[t!]
\centering
        \includegraphics[width=0.95\columnwidth,clip, angle=-90]{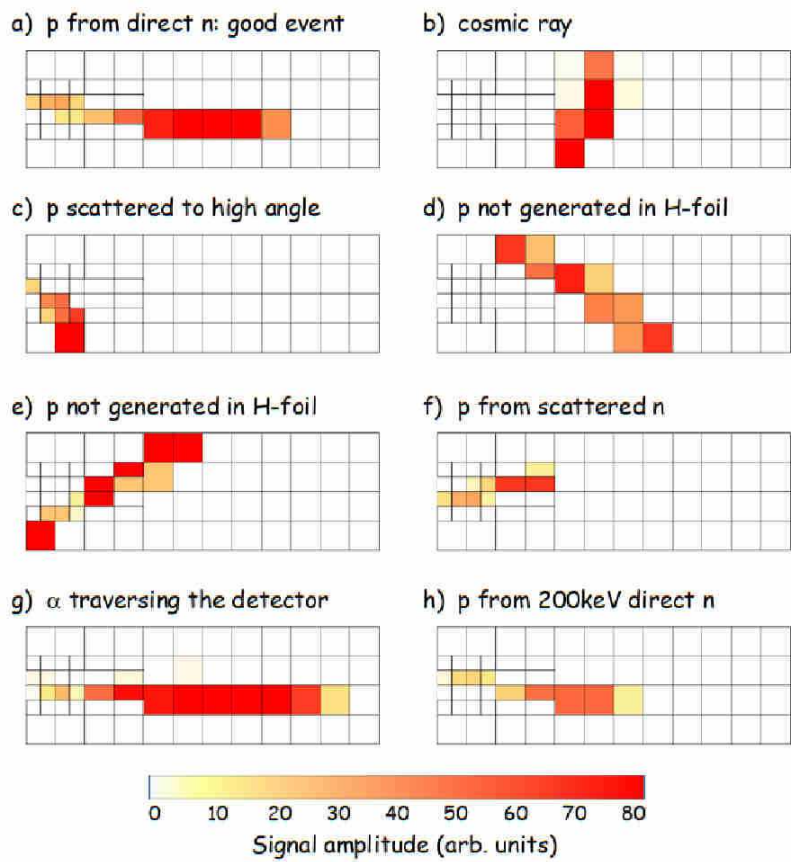} 
      \caption{Examples of reconstructed trajectories. Trajectories correspond to the detection of a) a proton generated from a direct neutron, i.e. a good event; b) a cosmic ray; c) a proton scattered to high angle; d-e) protons generated elsewhere than in the H-foil traversing (e) or not (d) the central collimator; f) proton generated by a scattered neutron; g) an $\alpha$ particle from the $\alpha$ source; f) a proton generated by a $200\,$keV direct neutron.}
\label{fig:detection plane and traces}
   \end{figure}
\paragraph{Proton track capabilities}
Examples of detected tracks are shown in Fig.\ref{fig:detection plane and traces}. 
Tracks are detected with sufficient precision to allow for faulty events rejection.
A first event selection is obtained 
imposing a $\Delta$E-E coincidence. This allows one to select potentially good events, as the one shown in Fig.\ref{fig:detection plane and traces}a, e and f, and to reject:
\begin{itemize}
\item[i] random events, like cosmics rays (Fig.\ref{fig:detection plane and traces}b).
\item[ii] recoil protons from the $^{1}$H foil scattered to high angles and therefore not going through the second collimator ($\Delta$E side only, Fig.\ref{fig:detection plane and traces}c)
\item[iii]  and protons generated elsewhere than in the $^{1}$H foil ($\Delta$E or E side only, Fig.\ref{fig:detection plane and traces}d) 
\end{itemize}
\begin{figure}[t]
\centering
\includegraphics[width=0.95\columnwidth,clip]{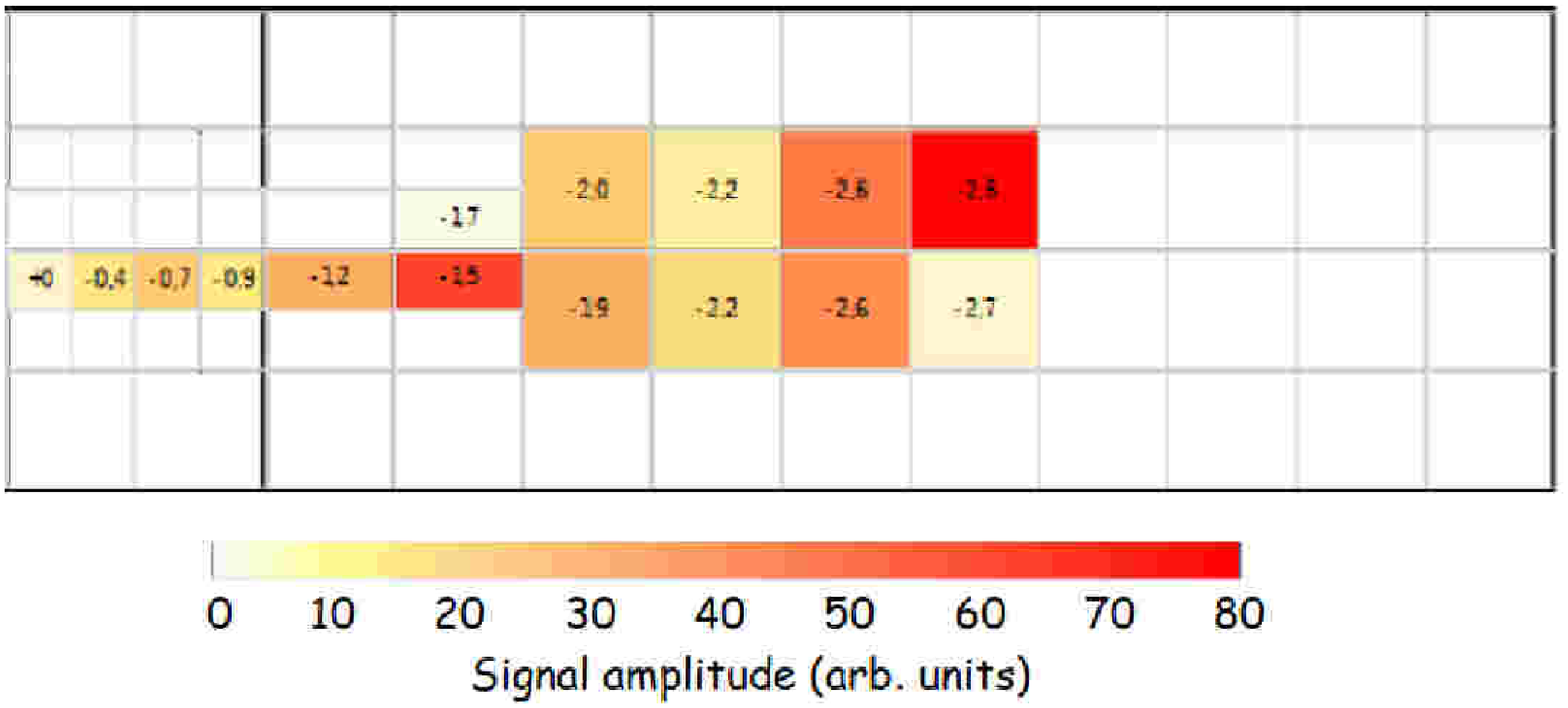} 
\caption{Tridimensional track reconstruction for  protons generated by a $1\,$MeV neutron. The values on the  track  are expressed in cm and represents the distance of the ionising track from the cathode.}
\label{fig:tracks3D}
\end{figure}
Moreover the  trajectory reconstruction allows the additional discrimination
\begin{itemize}
\item of recoil protons generated elsewhere than in the $^{1}$H foil but traversing the two collimators (Fig.\ref{fig:detection plane and traces}e)
\item  between direct (Fig.\ref{fig:detection plane and traces}a) and scattered neutrons (Fig.\ref{fig:detection plane and traces}f) impinging on the H-rich foil. In particular, protons knocked-out by scattered neutrons (which have lower energies than direct neutrons) have a shorter range in the gas and can be, to a certain extent, rejected.
\end{itemize}

A further improvement in the discrimination was obtained with a tridimensional trajectory reconstruction. When using the N$_{2}-$CO$_{2}$ gas, which has a slightly lower mobility than CF$_{4}$, it is possible to measure the drift time of  electrons towards each pad. This gives  access to the third coordinate of the trajectory, making the discrimination of ``good'' protons even more powerful.
 An example of such a 3D trajectory reconstruction is shown in Fig.\ref{fig:tracks3D}, where the numbers given on the pads  indicate the distance (in cm) of the ionising track from the cathode.

\paragraph{Relative energy calibration of each detection pad}
Scattered neutrons are rejected on the basis of their range. 
We stress here that the absolute energy deposited on each pad cannot be directly exploited because 
the gain of each detection pad is strongly dependent on the gas pressure, on the applied voltage, and on the local Micromegas structure, which can slightly vary from one pad to an other. 
The proton trajectory must however be reconstructed from the energy deposited on each pad, requiring a relative energy calibration.
The latter was obtained, for a given pressure and voltage, by analysing the signal amplitudes induced by an $\alpha$-particle traversing the detector (Fig.\ref{fig:detection plane and traces}g). Indeed, since the $\alpha$-Bragg peak is located at distances up to $10$ times the detector length, depending on the gas pressure, the energy deposited, and therefore the number of primary electrons on each pad, is about the same. Tracks 
 not spreading on several pads of the same raw were chosen and  measurements showed very similar gains for all the pads. 
\paragraph{Detector response in high neutron density environments}
The sensitivity of the detector to e$^{-}$ and/or $\gamma$-rays, generated by the neutron source, was investigated. 
In Ref.\cite{MariniNIMNoise}, we showed that Compton electrons generated from $\simeq 1\,$MeV $\gamma$-rays may give a significant signal 
in silicon detectors for mainly three reasons: i) they deposit their energy in the whole thickness of the detector, contrary to protons, which are stopped in few $\mu$m ($200\,$keV proton range in silicon is $2\,\mu$m)
; ii) the traversed thickness is greater than the detector thickness because they are mainly produced from $\gamma$ Compton scattering on surrounding materials and thus enter the detector with a significant angle, and because their slowing down is characterized by a more important angular straggling than for heavier charged particles; 
iii) given the high density of electrons, pile-up phenomenon may take place.
In the GPRT, these  processes are reduced by the low density of the gas and the adaptation of the gas pressure to the proton range. This is confirmed by the fact that no signal associated to electrons and $\gamma$-rays were detected during the tests.

We also remarked the absence of radiation damage signals induced by direct irradiation of the Micromegas and of the electronics with neutrons and/or $\alpha$ particles from the calibration source, thus confirming the possibility of using the GPRT  in a high neutron density environment (a few $10^{5}\,$n/s/cm$^{2}$).

\paragraph{Low-energy detection limit}
The low energy detection threshold of the detector was explored during the tests.
The detection of protons generated by down to $200\,$keV neutrons relies on i) the use of $1\,\mu$m $^{1}$H deposit foil; ii) the possibility of working at low gas pressure (below $30\,$mbar), where sparks may appear in the detector. 
To work with significantly polarized small gaps separated by low-pressure gas 
presents some technical difficulties.
Indeed, according to the Paschen's law, the breakdown voltage for a CF$_{4}$ pressure of $30\,$mbar and a gap in the Micromegas of $125\,\mu$m is slightly below $-450\,$V \cite{skoro2012}. Similarly, at this pressure, the breakdown voltage across the region where the cathode is  at less then $1\,$cm from grounded pieces, in particular from the rotation axis of the sample disk, is  $-1350\,$V.
To work with (p,V) values close to the breakdown voltage requires therefore a good control over the gas pressure in the detector, i.e. reduced  fluctuations. 
Tests were performed at a CF$_{4}$ pressure of $40\,$mbar and with an $^{1}$H foil thickness of $4\,\mu$m. Despite the non-optimal experimental conditions, traces of the most energetic protons generated by $300\,$keV neutrons could be observed.
When using N$_{2}-$CO$_{2}$, proton traces at neutron energies as low as $200\,$keV could be obtained at a gas pressure of $30\,$mbar, as shown in Fig.\ref{fig:detection plane and traces}h.
\subsection{Electrostatic behavior and field cage} \label{sec:electrostatic behavior}
\begin{figure}[ht!]
\centering
        \includegraphics[width=0.95\columnwidth,clip]{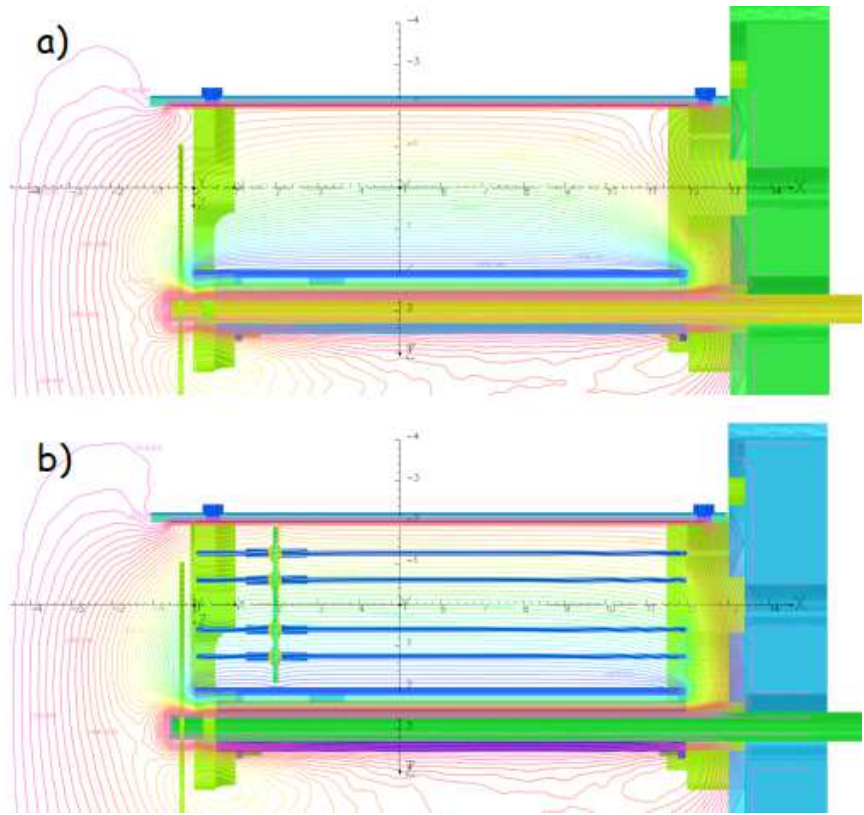}
      \caption{Equipotential lines simulated with the OPERA code. The field cage is present (b) and has been removed from the simulated setup (a). 
      The potentials applied to the cathode and to the Micromegas gap are of -$2000\,$V and -$500\,$V, respectively.
}
\label{fig:LignesEquip NoCageADerive NoChargeSurCollimateur}
   \end{figure}

From an electrostatic point of view, a fundamental requirement for the detector to properly work is the uniformity of the electric field between the electrodes.
\paragraph{Non-uniformity of the electric field}
The finite dimensions of the detector and the presence of surrounding conducting materials (rotation axis of the sample disk, guard-ring 
of Micromegas, reaction chamber...) distort the electric field between the electrodes. 
To a less extent, the electric field is distorted also by the presence of insulators (Macor collimators) in the space between the electrodes. 
\begin{figure}[t!]
\centering
        \includegraphics[width=0.95\columnwidth,clip]{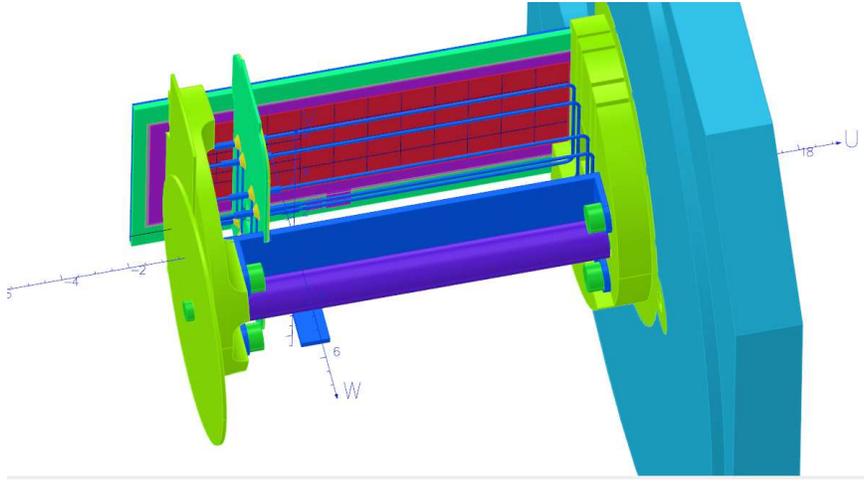} 
      \caption{Detector geometry implemented in OPERA simulations \cite{OPERA}.}
\label{fig:OPERA whole det}
   \end{figure}
Electrostatics simulations of the whole detector were performed with the code OPERA \cite{OPERA}. The implemented geometry is shown in Fig.\ref{fig:OPERA whole det}.
In Fig.\ref{fig:LignesEquip NoCageADerive NoChargeSurCollimateur}(a)  the equipotential lines obtained applying -$2000\,$V and -$500\,$V  to the cathode and to the Micromegas, respectively, are presented. The field is distorted at the entrance and at the exit of the detector, due to the presence of grounded structures. 
First experimental tests of the GPRT prototype without field cage confirmed these results, showing a bad charge collection for the external pads of the detection plane.
Simulations with a field cage show a better uniformity of the electric field (Fig.\ref{fig:LignesEquip NoCageADerive NoChargeSurCollimateur}(b)), although small deviations are still observed at the back of the detector.

\begin{figure}[t!]
   \centering
      \includegraphics[width=0.95\columnwidth,clip]{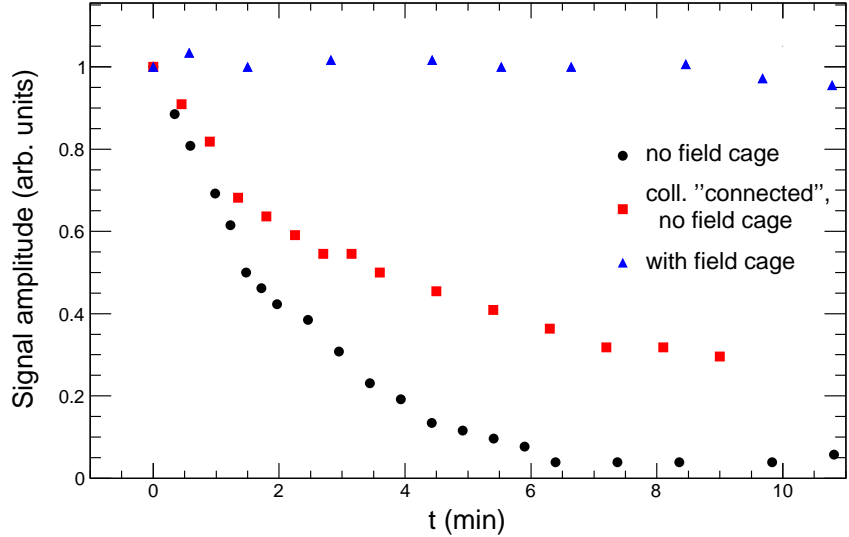}
      \caption{Evolution of the signal amplitude of the first raw of pads with irradiation time $t$ without (circles and squares) and with the field cage (triangles). Squares indicate the setup where the entrance collimator was grounded and connected to the cathode. For comparison, signal amplitudes are normalised to 1 at the beginning of the irradiation.}
\label{fig:charge accumulation}       
  \end{figure}

\begin{figure}[t!]
 \centering
        \includegraphics[width=0.95\columnwidth,clip]{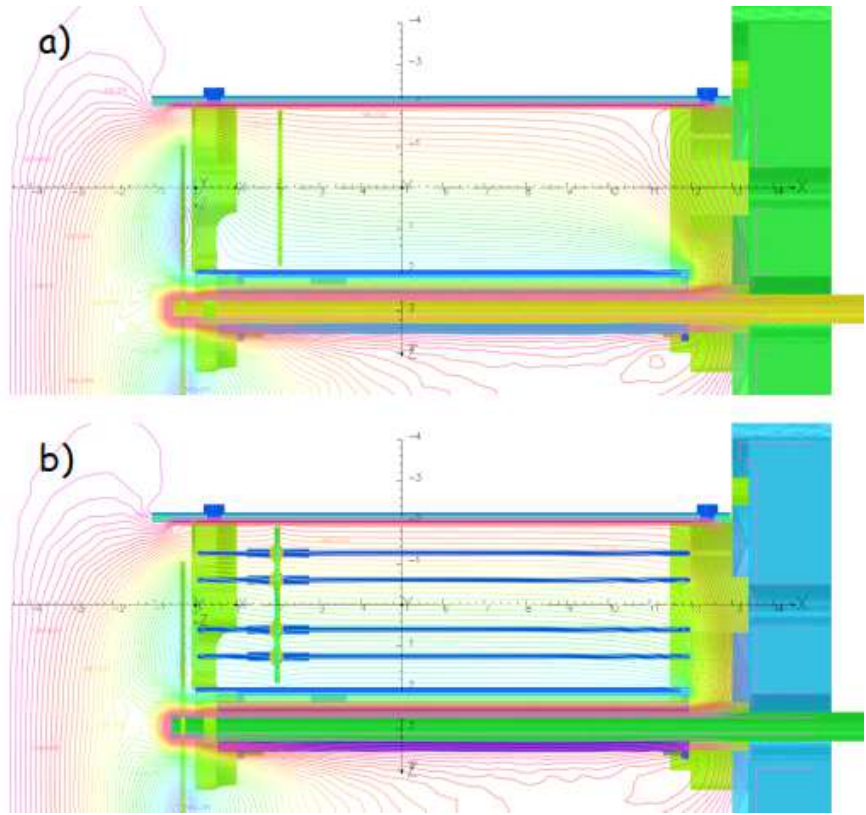} 
      \caption{As Fig.\ref{fig:LignesEquip NoCageADerive NoChargeSurCollimateur}
     The field cage is present (b) and has been removed from the simulated setup (a). A static charge of $-2\,$nC is added on the entrance collimator.
     }
\label{fig:LignesEquip NoCageADerive ChargeSurCollimateur}
   \end{figure}

\paragraph{Electrostatic charges accumulation}
During tests without the field cage, it was also observed that, after few minutes of irradiation, the signal amplitude of the first raw of pads
decreases of about $95\%$, as shown in Fig.\ref{fig:charge accumulation} (circles).
The reduction of the charge collection efficiency was attributed to a charge accumulation on the insulator pieces under beam irradiation, and in particular on the entrance collimator.
As already discussed, the neutron production is associated to the generation of an elevated number of electrons at different energies \cite{MariniNIMNoise}. 
These electrons contribute to the polarization of the sample disk and the entrance collimator, progressively modifying the electric field at the entrance of the detector.
 Therefore primary electrons generated at the entrance of the chamber progressively drift towards inner pads, 
inducing a progressive reduction of the signal amplitude on the pads closest to the insulator pieces.
 The electrostatic charge could be partially removed and one third of the full signal amplitude kept by connecting the entrance collimator to the cathode and grounding it (squares in Fig.\ref{fig:charge accumulation}). 
  This effect has been investigated in OPERA simulations.
A static charge of $-2\,$nC was added to the sample disk
 in the simulation. 
 It should be noted that, being the accumulated electric charge very difficult to estimate, the value of $-2\,$nC in the simulation was chosen on the basis of the observed effects and the simulation is only  qualitative. 
 The distortion of the electric field is shown in Figs.\ref{fig:LignesEquip NoCageADerive ChargeSurCollimateur}(a) and (b), without and with a field cage, respectively. It should be remarked that, without a field cage, electrons generated at the entrance of the detector cannot reach the first pads, as it was observed during the tests, and are driven back onto the first collimator. The presence of a field cage significantly reduces this effect.
Experimental tests carried out with the field cage have shown a good stability of the signal amplitude as function of the irradiation time, as shown in Fig.\ref{fig:charge accumulation} (triangles), indicating a reduction of the electric field distortion, which does not significantly affect the charge collection.

\section{Conclusions}\label{Conclusions}
Accurate neutron flux measurements via the elastic scattering $^{1}$H(n,p) reaction is challenging for neutron energies below $1\,$MeV, because of the high background contribution, mostly due to gamma-rays/electrons produced in the neutron production target. To overcome this issue, a new gaseous proton recoil telescope constituted of a  double ionization chamber was designed and experimentally tested at the CENBG with monoenergetic neutrons. The detector has a large dynamical range thanks to its adjustable gas pressure and sample thickness, and the use of Micromegas technology. 
Its low electron sensitivity makes it well suited for neutron energies down to $200\,$keV in a $\gamma$s and electrons dense environment.
The $\Delta$E-E coincidences, coupled to a tridimensional proton track
analysis, enables the disentanglement between direct neutrons
impinging on the PP foil, scattered neutrons impinging on the PP foil,
and direct neutrons impinging on other H-rich surrounding materials.
Electrostatic simulations showed that the electric field is slightly distorted by surrounding
materials, and heavily distorted by the space charge accumulation on
ceramic pieces of the detector.
Experimental tests, supported by simulations, indicate that the use of a field cage allows to overcome both issues.
The experimental results obtained so far indicate that the GPR telescope
can be useful in several applications where a monoenergetic neutron flux extending from few MeV down to $200\,$keV needs to be precisely measured.
A measurement of the proton detection efficiency of the detector, and in particular that every acceptable recoiling proton is actually detected, is planned for the second half of 2019 at the AIFIRA facility and will require a quantitative experiment, with 
the use of a microbeam delivering a well known number of protons.
\section*{Acknowledgments}
We wish to acknowledge the support of the AIFIRA staff for providing the beam and of CEA Saclay for providing the Micromegas\\
This work was supported by the European Commission within the EURATOM FP7 2007-2013 framework program through CHANDA  (project No. FP7-605203). 
\section*{References}
\bibliographystyle{unsrt}
\bibliography{bibliogr}

\end{document}